\documentclass[prl,preprint,showpacs]{revtex4}
\usepackage{graphicx}
\usepackage{amssymb}
\usepackage{bm}
\usepackage{color}
\begin{document}
\newcommand{\be}{\begin{equation}}
\newcommand{\ee}{\end{equation}}
\newcommand{\bea}{\begin{eqnarray}}
\newcommand{\eea}{\end{eqnarray}}

\setcounter{page}{1}

\title[]{Yang-Lee Zeros of the Triangular Ising Antiferromagnets}

\author{Chi-Ok \surname{Hwang}}
\affiliation{Division of Computational Sciences in Mathematics, National Institute 
for Mathematical Sciences, Daejeon 305-340, Korea
}

\author{Seung-Yeon \surname{Kim}}
\affiliation{School of Liberal Arts and Sciences, Chungju National University,
Chungju, 380-702, Korea
}

\date{\today}

\begin{abstract}
 Using both the exact enumeration method (microcanonical transfer matrix) for a small system ($L=9$) 
and the Wang-Landau Monte Carlo algorithm for large systems to $L=30$, we obtain the exact and 
approximate densities of states g(M,E), as a function of magnetization M and exchange energy E, 
for the triangular-lattice Ising model. Based on the density of states $g(M,E)$, 
we investigate the phase transition properties of Yang-Lee zeros for the triangular Ising
antiferromagnets and obtain the magnetic exponents at various temperatures.
\end{abstract}
\maketitle

In 1952, it was proposed by Yang and Lee a new theory for explaining the occurrence 
of phase transitions 
in the thermodynamic limit~\cite{PR_CNYang1952,PR_TDLee1952}. 
They reinterpreted the partition function as a polynomial of the exponential variable
including magnetic field $\beta H$. They proposed that in the thermodynamic limit the real axis cross 
of the complex zero set of the polynomial is directly related to the phase transition. 
They illustrated their approach by solving the lattice gas (ferromagnetic Ising model 
in a magnetic field) problem exactly. Later on, along with computational developments 
this approach has been extended to treating other exponential variables 
like the exponential term including temperature by Fisher and others~\cite{MEFisher1965}. 
In various applications, computational improvements enabled researchers to get 
the exact or approximate density of states 
(DOS) of the finite systems~\cite{PRE_CZhou2005,PRL_SYKim2004,AJP_DPLandau2004,PRE_Schulz2003,
PRE_Wang2001,PRL_WJanke1995,PRL_JLee1993,PRL_BABerg1992,PRL_UWolff1989,PRL_RHSwendsen1987}. 
However, the extraction of the density of zeros for a finite and numerically accessible lattice
sizes had been considered very challenging. In recent years, there have been some attempts to overcome
the difficulties~\cite{CPC_WJanke2002}. 

Exact DOS can be obtained only for very finite systems like up to the
linear size $L=9$ for the
bipartite system on the two dimensional (2D) triangular lattice with nearest neighbour 
interactions~\cite{JSM_COHwang2007,NPB_SYKim2008}. However, approximate methods like the Wang-Landau 
sampling~\cite{PRE_CZhou2005,AJP_DPLandau2004,PRE_Schulz2003,PRE_Wang2001}
can obtain DOS of the quite large finite systems, for example, up to
$L=30$ for the bipartite system on the 2D triangular lattices with nearest 
neighbour interactions~\cite{JSM_COHwang2007}. 

In this paper, as a series of investigation for the 2D triangular lattice systems
with nearest neighbour interactions~\cite{JSM_COHwang2007,JKPS_COHwang2008,NPB_SYKim2008} 
Yang-Lee zeros are investigated.
We construct a high-degree polynomial to get Yang-Lee zero set 
using the density of states from~\cite{JSM_COHwang2007}. 
In this case, the Hamiltonian ${\cal H}$ is given as follows;
\begin{equation}
{\cal H}=-J\sum_{i,j}\sigma_i \sigma_j - H \sum_i \sigma_i,
\end{equation}
where $E=\sum\limits_{\langle i,j\rangle} \sigma_i \sigma_j$ is the
exchange energy, $M=\sum\limits_{i=1}^N \sigma_i$ the total
magnetization, $J$ the coupling constant ($J > 0$ for ferromagnets (FM)
and $J < 0$ for antiferromagnets (AFM): In this paper, for simplicity we take $J=-1$.), $\langle i,j\rangle$ denotes
distinct pairs of nearest neighbor sites, $H$ the external magnetic field and $\sigma_i=\pm 1$.
The polynomial is
\begin{equation}
Z(a,x)=\sum_E [\sum_M g(M,E) a^E] x^M.
\end{equation}
Here, $a=e^{-2\beta}$, $x=e^{-2h}$ and the reduced magnetic field $h=\beta H$. 

\begin{figure}
\rotatebox{-90}{
\includegraphics[width=4.1cm]{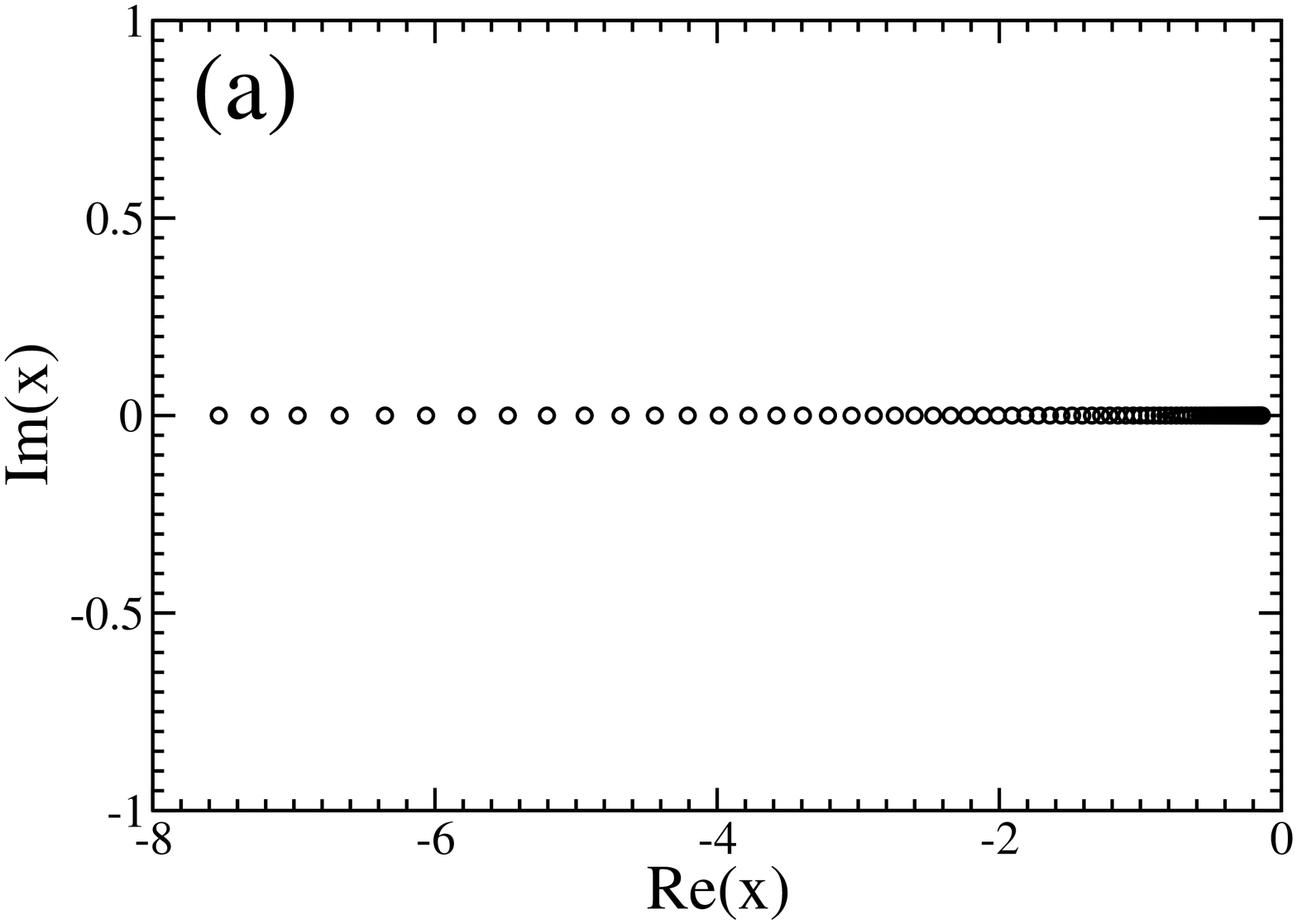}}
\rotatebox{-90}{
\includegraphics[width=4.1cm]{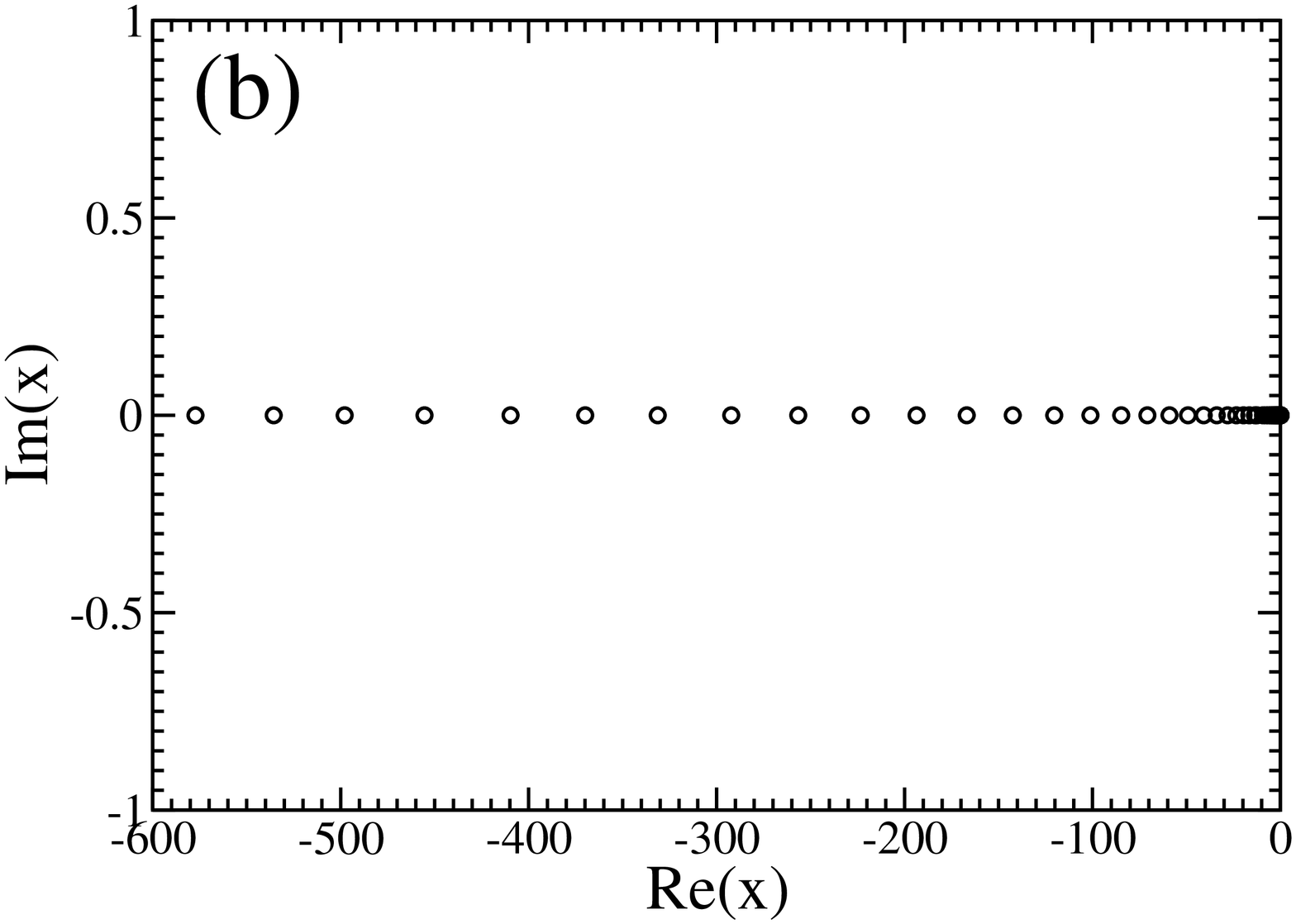}}
\rotatebox{-90}{
\includegraphics[width=4.1cm]{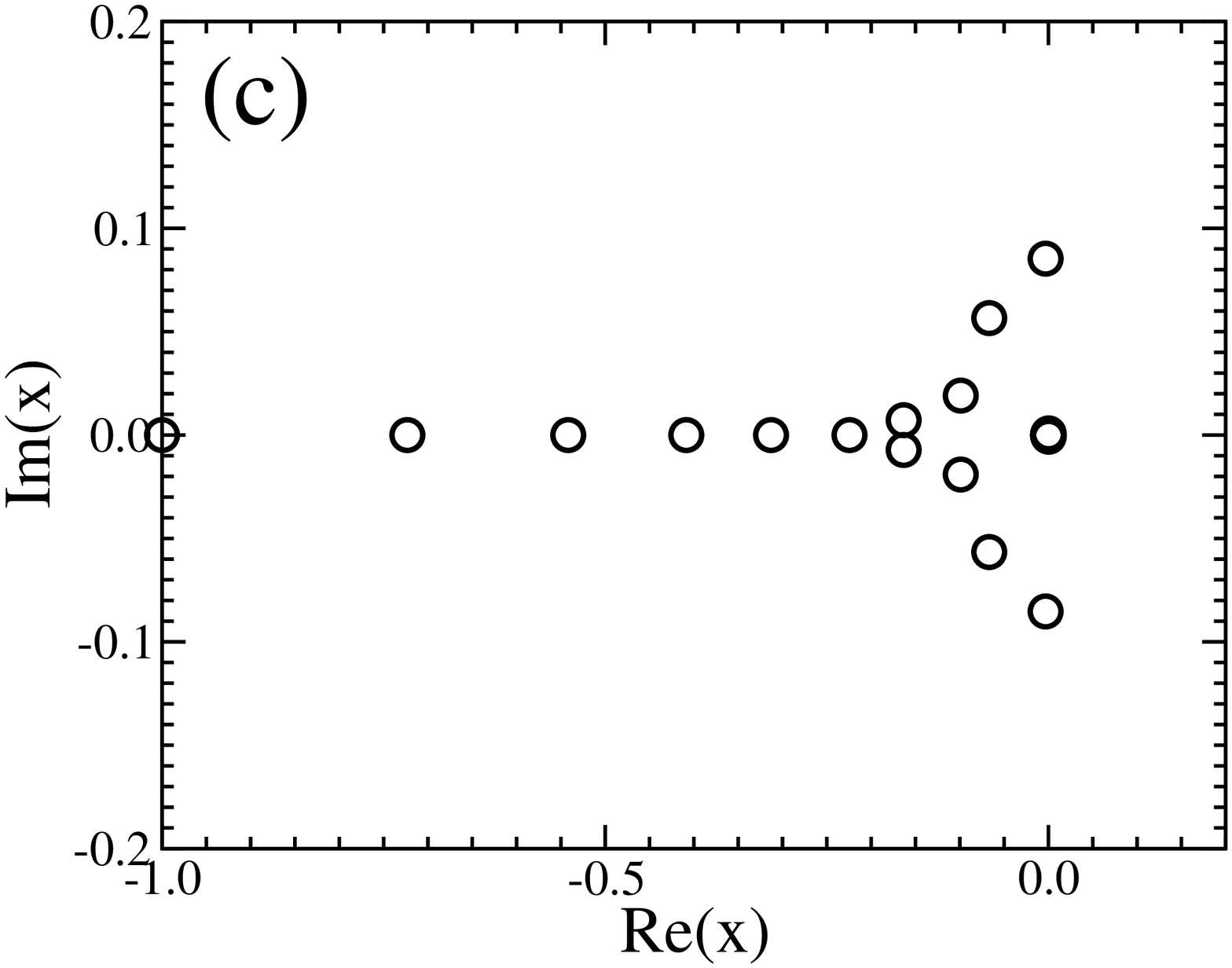}}
\caption{Yang-Lee zeros in the complex $x$ plane of the $9 \times 9$ triangular AF Ising model with
periodic boundary conditions: (a) $a=0.9$, (b) $a=0.5$, and (c) $a=0.2$.
}
\label{Yang-Lee_a}
\end{figure}

\begin{figure}
\rotatebox{-90}{
\includegraphics[width=7cm]{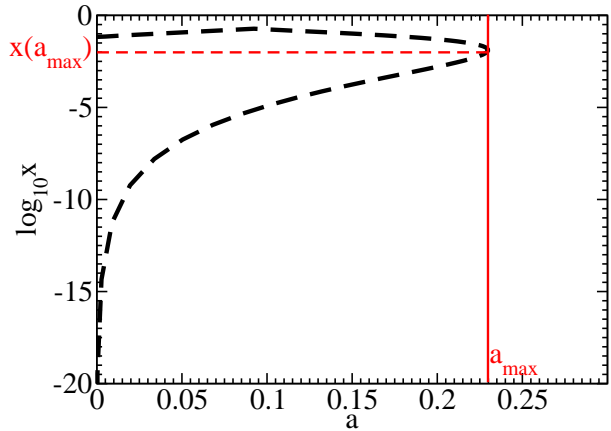}}
\caption{Typical phase diagram (black dotted line) of the triangular AF Ising
model~\cite{JKPS_COHwang2008}. Below $a_{max}\approx 0.23$ (red solid line),
for a given critical temperature $a$ there are two critical magnetic fields.
The critical value of $x(a_{max})$, corresponding to $a_{max}\approx 0.23$,
is approximately $0.012$.}
\label{CTHDiagram}
\end{figure}

At first, using the exact density of states (for $L=9$) we investigate the Yang-Lee zeros 
in the complex $x$ plane at several different temperatures, $a=0.2$, 0.5 and 0.9.
Figure~\ref{Yang-Lee_a} shows the Yang-Lee Zeros in the complex $x$ plane
of the $9 \times 9$ AF Ising model with the periodic boundary conditions.
At high temperatures like $a=0.9$ and 0.5, all zeros lie in the left half plane
(Fig.~\ref{Yang-Lee_a} (a) and (b)).
As the temperature decreases, some of them move toward the right half plane.
At low enough temperatures
like Fig.~\ref{Yang-Lee_a} (c),
where there are phase transitions (see Fig.~\ref{CTHDiagram}),
the Yang-Lee zeros begin to appear in the right half plane.

It is well known the phase diagram of the triangular antiferromagnets in temperature-magnetic 
field plane~\cite{PL_BDMetcalf1973,PRB_MSchick1977,PRB_WKinzel1981,IJMP_JDNoh1992,JKPS_COHwang2008}.
In the phase diagram, it is noted that there are two critical magnetic fields
for a given critical temperature because of the critical line shape of the phase diagram.
Figure~\ref{CTHDiagram} shows the typical phase diagram in reduced temperature-magnetic
field plane. We use the terms, \lq\lq low\rq\rq ($x > 0.012$) and 
\lq\lq high\rq\rq ($x < 0.012$) reduced magnetic fields.

\begin{figure}
\rotatebox{-90}{\includegraphics[width=6cm]{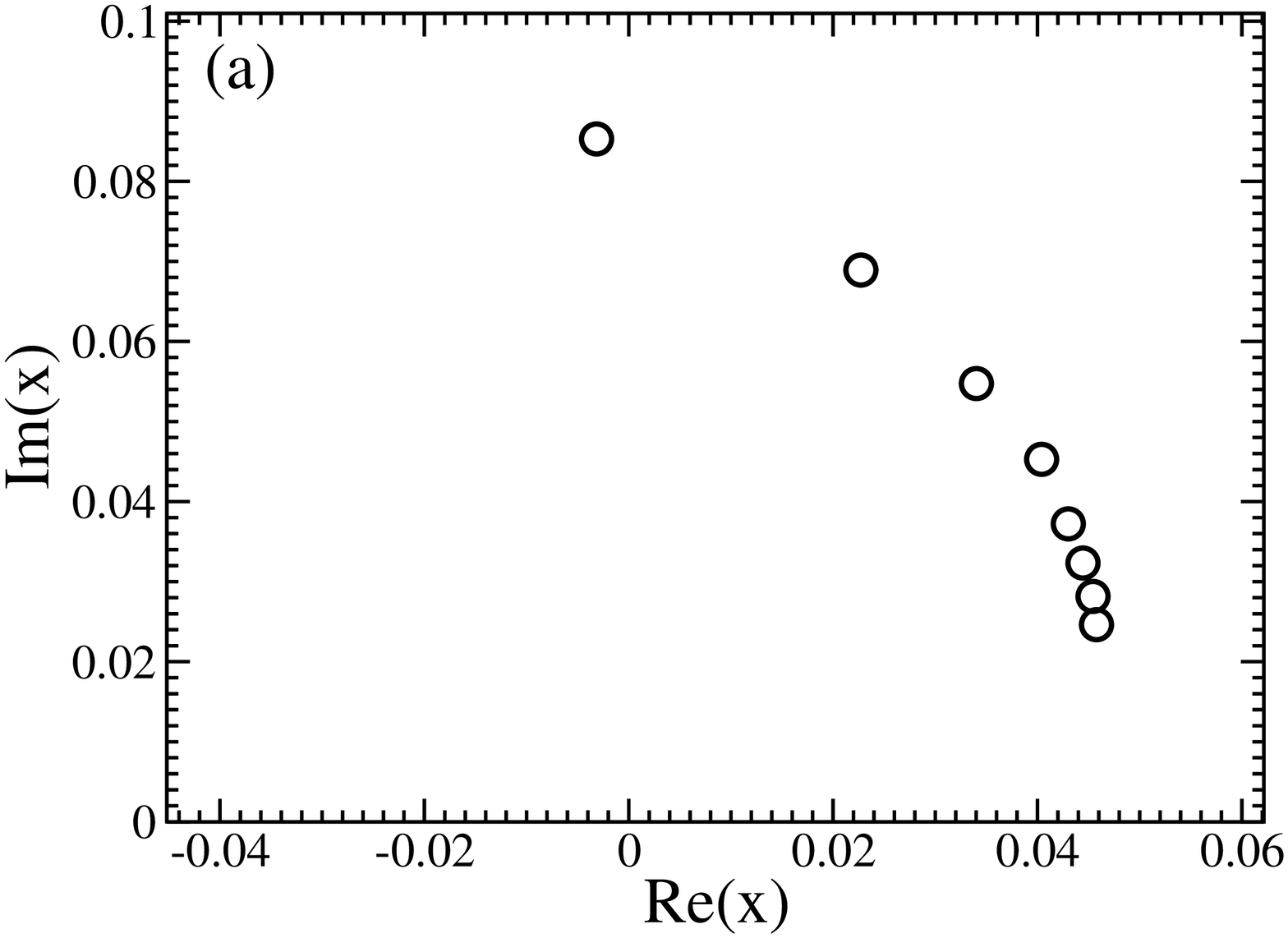}}
\rotatebox{-90}{\includegraphics[width=6cm]{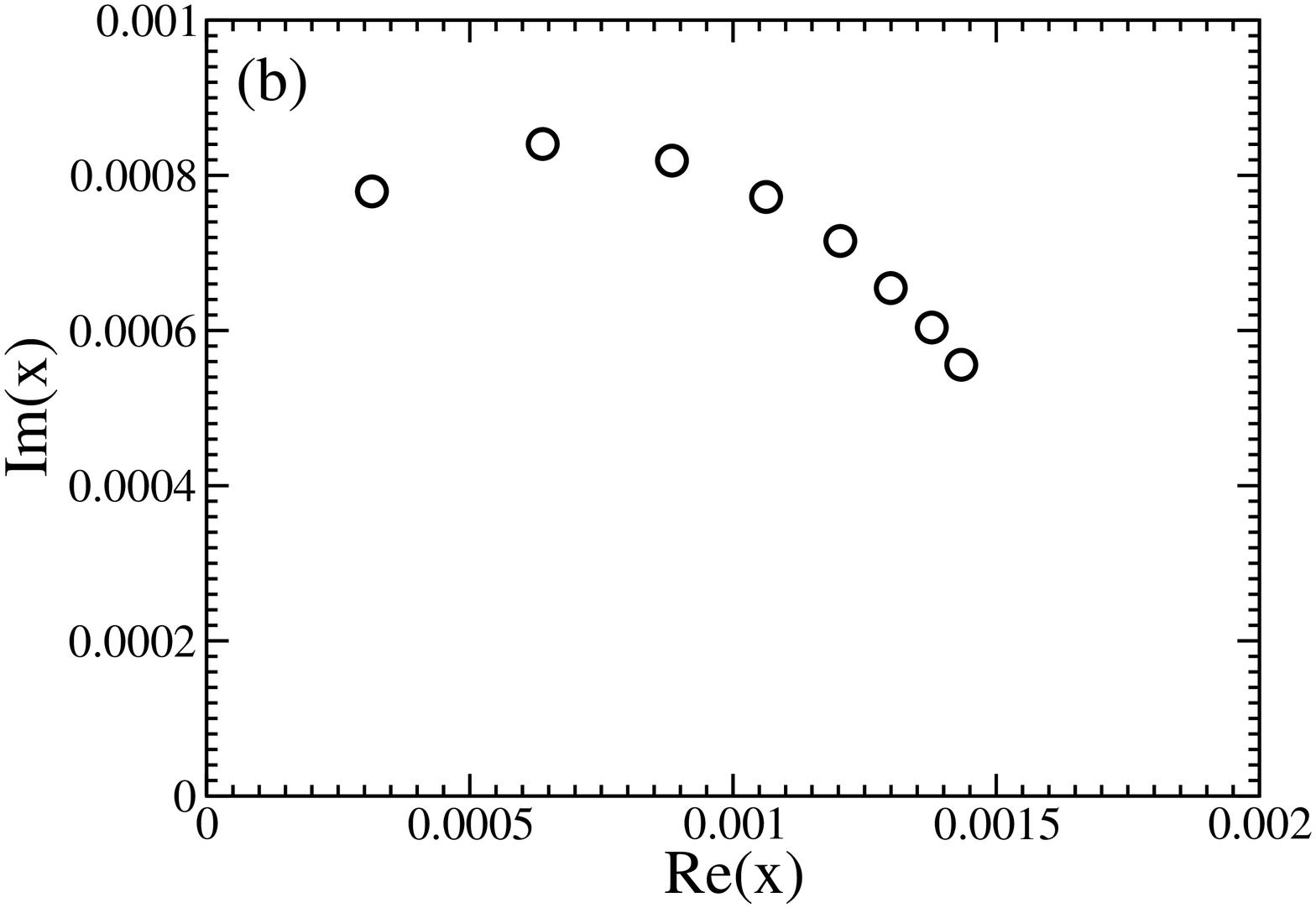}}
\caption{First Yang-Lee zeros of the various system sizes, $L=9-30$ (three multiples) 
(a) in low and (b) high magnetic fields at $a=0.2$:
}
\label{Yang-Lee0.2}
\end{figure}

\begin{figure}
\rotatebox{-90}{
\includegraphics[width=6.0cm]{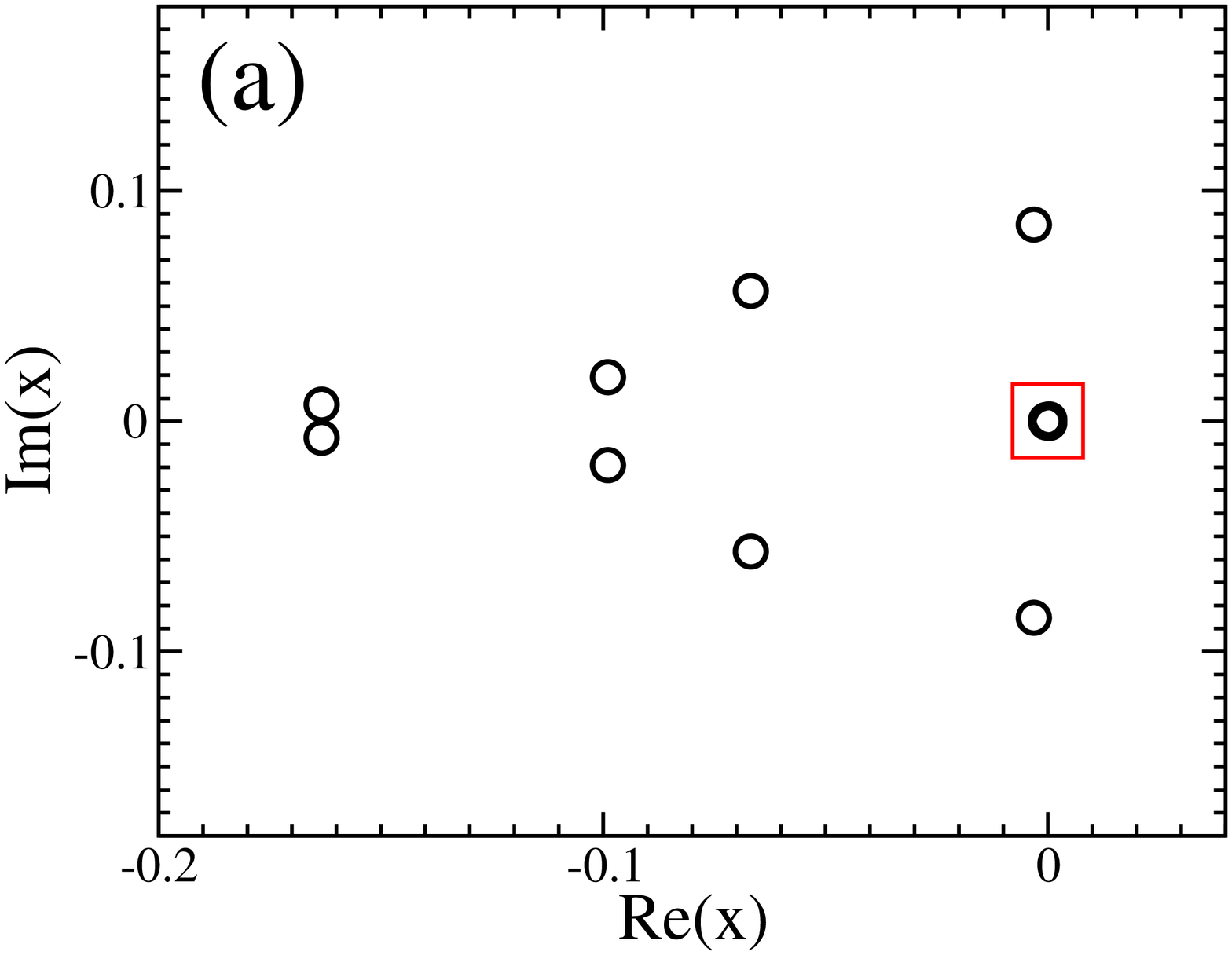}}
\rotatebox{-90}{
\includegraphics[width=6.0cm]{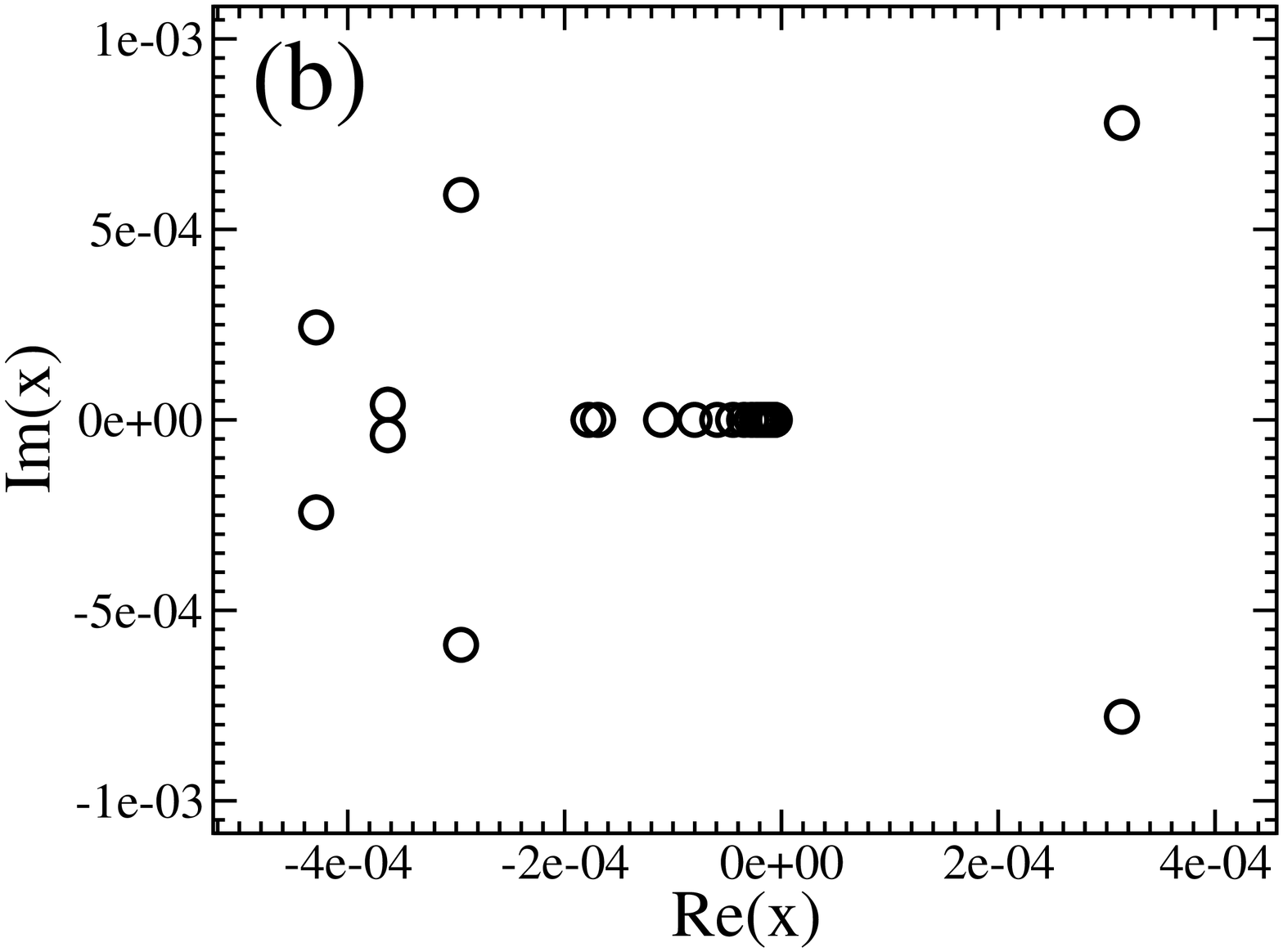}}
\caption{Figures of Yang-Lee zeros in the complex $x$ plane of the $9 \times 9$
triangular AF Ising model with
periodic boundary conditions at $a=0.2$ : (a) in low magnetic field ($Re(x) \sim 0.01$),
and (b) in high magnetic field ($Re(x) \ll 0.01$), zoomed-in figure around the origin of
the left figure (boxed area).
}
\label{Yang-Lee_a0.2}
\end{figure}

For several temperatures, $a=0.05$, 0.1, 0.15 and 0.2, we obtain the Yang-Lee
zeros for different system sizes (three multiples from 9 to 30) and investigate
whether there is the real axis cross of the first Yang-Lee zeros in the thermodynamic limit.
In an illustrated Fig.~\ref{Yang-Lee0.2}, we show the first Yang-Lee zeros 
for the various system sizes
in high (\ref{Yang-Lee0.2} (a)) and low (\ref{Yang-Lee0.2} (b)) magnetic fields 
at $a=0.2$. In the figure, it should be observed that the two graphs (a)
and (b) are in completely different scales. For a given temperature $a$,
the Yang-Lee zeros for low and high magnetic fields are in completely different scales 
in one Yang-Lee zero solution plane. As a typical example in Fig.~\ref{Yang-Lee_a0.2},
the Yang-Lee zeros are shown in high and low magnetic fields.

Table~\ref{table} illustrates the real and imaginary parts of the first zeros $x_1$ at $a=0.2$ 
in the high magnetic field for
$L=9-30$ (three multiples). Using the Bulirsch-Stoer algorithm~\cite{PRL_SYKim2004,JPA_Henkel1988}, 
we extrapolated our results for the finite 
lattices to infinite size and obtained $x_1= 0.0019(5) - 0.00001(9) \it{i}$, 
indicating the phase
transition of the AF Ising model in an external magnetic field. Note that we use 
approximate density of states from Monte Carlo results
for large systems ($L=12-30$, three multiples) so that there are some errors 
even in the first Yang-Lee zeros. For example, in our data (second column in Table~\ref{table_Yh}), 
the imaginary part of the first zero at infinite size for $a=0.05$ (low) is positive with much error,
we believe, due to the well-known strong crossover effects~\cite{PRB_HWJBlote1993}
 for low magnetic fields. 
In thermodynamic limit, if there is a phase transition, 
the imaginary part of the first zero should be zero
if we use exact first zeros for the extrapolation to infinite size and the extrapolation
is exact. However, it should be also mentioned that overall the first zeros of 
approximate density of states from Monte Carlo results are very reliable (for example, 
in the triangular lattices for $L=9$, the first zeros of exact density of states 
at the first line in Table~\ref{table} are exactly the same to the first zeros of 
approximate density of states from Monte Carlo results).

\begin{table}[t]
\begin{center}
\caption{Real and imaginary parts of the first zeros $x_1$ at $a=0.2$ for $L = 9-30$
(three multiples)
in the high magnetic field. $y_h(L)$ is the scaling exponent calculated
by Eq.~\ref{y_h}. The last row is
the extrapolation to infinite size.}
\begin{tabular}{r c c l} \hline \hline
$\qquad L\qquad$ & Re($x_1$) & Im($x_1$) & $y_h(L)$    \\ \hline
$\qquad 9\qquad$    &   0.0003141075  &  0.0007793222  &  -0.262 \\
$\qquad 12\qquad$   &   0.0006385131  &  0.0008404157  &  0.115  \\
$\qquad 15\qquad$   &   0.0008840487  &  0.0008190623  &  0.323  \\
$\qquad 18\qquad$   &   0.0010628038  &  0.0007721926  &  0.495  \\
$\qquad 21\qquad$   &   0.0012038568  &  0.0007154301  &  0.663  \\
$\qquad 24\qquad$   &   0.0012996089  &  0.0006547772  &  0.687  \\
$\qquad 27\qquad$   &   0.0013774157  &  0.0006038531  &  0.787  \\
$\qquad 30\qquad$   &   0.0014334636  &  0.0005557930  &         \\
$\qquad \infty \qquad$ &     $\qquad 0.0019(5)$ $\qquad$ & $\qquad -0.00001(9)$ $\qquad$ & 1.297(28) \\ \hline \hline
\end{tabular}
\label{table}
\end{center}
\end{table}

Also, we can obtain the magnetic scaling exponent 
for the finite linear size $L$~\cite{NCB_CItzykson1983}
\begin{equation}
y_h(L) = - \frac{\ln\{{\rm Im}[x_1(L+3)]/{\rm Im}[x_1(L)]\}}{\ln[(L+3)/L]}, \label{y_h}
\end{equation}
where $x_1$ is the first zero. The fourth column of Table~\ref{table} shows the values of the scaling
magnetic exponent $y_h(L)$. The extrapolated value is $y_h=1.30$ at $a=0.2$ 
in the high magnetic field.
Similarly, in the thermodynamic limit we have obtained $x_c$ and $y_h$ for $a=0.05$, 0.1, 0.15
and 0.2. They are tabulated in Table~\ref{table_Yh}. The evaluated $y_h$ ranges 
from $1.3$ to $1.7$ depending on the temperature $a$~\cite{PRE_SLAQueiroz1999}.
It is known that for $h=6$ and $T=0$ the triangular antiferromagnets map onto Baxter's hard-hexagon
lattice  gas of which the critical exponents are exactly known~\cite{PRB_WKinzel1981,PRE_XQian2004}.
In Table II, we observe that as the value $a$ decreases in high magnetic fields $y_h$ gets bigger
to $28/15$. The estimated $y_h$ values may indicate that the triangular Ising antiferromagnets belong
to the three-state Potts class for high magnetic fields but not to the class for intermediate 
(around $h=3$) and low magnetic fields. These results contradict with the previous 
results~\cite{IJMP_JDNoh1992,PRE_SLAQueiroz1999}, which say that the phase transition belongs
to the 3-state Potts universalty class over the whole phase boundary (that means that 
$y_h$ should be $28/15$ in the whole phase boundary). This might say that our results
reflect the well-known strong crossover effects~\cite{PRB_HWJBlote1993} 
for intermediate (around $h=3$) and low magnetic 
fields and slow convergence of $y_h$. More research is required on this matter. 

As a whole, from the imaginary parts of the second column in Table II, 
it is confirmed that there are phase transitions in external magnetci fields 
for the various temperatures.
 
\begin{table}[t]
\begin{center}
\caption{The critical points $x_c(a)$ and the magnetic scaling exponent $y_h$ \it{vs} \rm $a$, 
in the limit $L \to \infty$: \lq\lq low" 
and \lq\lq high"  represent low and high magnetic fields respectively for the same $a$.
Note that there is no phase transition for high temperatures, that is, large $a$ values 
(see Fig.~\ref{CTHDiagram}). }
\begin{tabular}{c  c l} \hline \hline
$\qquad a\qquad$ & $x_c$  & $y_h$    \\ \hline
$\qquad 0.05$ (low)$\qquad$  & 0.8(2) + 0.17(3)\it{i}   $\qquad$&  1.7(1) \\
$\qquad 0.1$ (low)$\qquad$  & 0.22(6) + 0.02(4)\it{i}    $\qquad$&  1.7(3)  \\
$\qquad 0.15$ (low)$\qquad$ & 0.102(7) + 0.1(1)\it{i}   $\qquad$ &  1.61(8)  \\
$\qquad 0.2$ (low)$\qquad$  & 0.046(4) - 0.4(4)\it{i}  $\qquad$ &   1.5(1)  \\
$\qquad 0.05$ (high)$\qquad$ &  1.83(3)e-7 + 0.0(1)e-7\it{i}    $\qquad$ &  1.44(1)  \\
$\qquad 0.1$ (high)$\qquad$ & 1.31(5)e-5 - 0.97(16)e-6\it{i}    $\qquad$&  1.38(4)  \\
$\qquad 0.15$ (high)$\qquad$&   1.9(1)e-4 - 0.0(4)e-4\it{i}    $\qquad$ &  1.49(2)  \\
$\qquad 0.2$ (high)$\qquad$ &  1.9(5)e-3 - 0.1(9)e-4\it{i} $\qquad$ &  1.297(28)      \\ \hline \hline
\end{tabular}
\label{table_Yh}
\end{center}
\end{table}

\newpage
\bibliographystyle{unsrt}
\bibliography{/home/chwang/Research_NIMS/Research_0/Bib_DB/Mine/Hwang}

\end{document}